\documentclass[preprint2,times,tighten]{aastex6}
\pdfoutput=1 %for arXiv submission
\usepackage{amsmath,amstext}
\usepackage[figure,figure*]{hypcap}
\usepackage{newtxmath} %use times font for math

\shorttitle{How do bulges and BHs grow?}
\shortauthors{Bell et al.}

\begin{document}

\title{Galaxies grow their bulges and black holes in diverse ways}
\author{Eric F.\ Bell\altaffilmark{1}, Antonela Monachesi\altaffilmark{2}, Benjamin Harmsen\altaffilmark{1}, Roelof S.\ de Jong\altaffilmark{3}, Jeremy Bailin\altaffilmark{4}, David J.\ Radburn-Smith\altaffilmark{5}, Richard D'Souza\altaffilmark{1}, Benne W.\ Holwerda\altaffilmark{6}
}
\email{ericbell@umich.edu}
\altaffiltext{1}{Department of Astronomy, University of Michigan, 1085 South University Ave., Ann Arbor, MI 48109-1107}
\altaffiltext{2}{Max Planck Institut f{\"u}r Astrophysik, Karl-Schwarzschild-Str. 1, Postfach 1317, 
D-85741 Garching, Germany}
\altaffiltext{3}{Leibniz-Institut f\"ur Astrophysik Potsdam (AIP),
An der Sternwarte 16, 14482 Potsdam, Germany}
\altaffiltext{4}{Department of Physics and Astronomy, University of Alabama, Box 870324, Tuscaloosa, AL 35487-0324, USA}
\altaffiltext{5}{Department of Astronomy, University of Washington, 3910 15th Ave NE, Seattle, WA 98195, USA}
\altaffiltext{6}{Department of Physics and Astronomy, University of Louisville, 102 Natural Science Building, Louisville, KY 40292, USA}
\begin{abstract}
Galaxies with Milky Way-like stellar masses have a wide range of bulge and black hole masses; in turn, these correlate with other properties such as star formation history. While many processes may drive bulge formation, major and minor mergers are expected to play a crucial role. 
Stellar halos offer a novel and robust measurement of galactic merger history; cosmologically-motivated models predict that mergers with larger satellites produce more massive, higher metallicity stellar halos, reproducing the recently-observed stellar halo metallicity--mass relation. We quantify the relationship between stellar halo mass and bulge or black hole prominence using a sample of eighteen Milky Way-mass galaxies with newly-available measurements of (or limits on) stellar halo properties. {\it There is an order of magnitude range in bulge mass, and two orders of magnitude in black hole mass, at a given stellar halo mass (or, equivalently, merger history).} Galaxies with low mass bulges show a wide range of quiet merger histories, implying formation mechanisms that do not require intense merging activity. Galaxies with massive `classical' bulges and central black holes also show a wide range of merger histories. While three of these galaxies have massive stellar halos consistent with a merger origin, two do not --- merging appears to have had little impact in making these two massive `classical' bulges. Such galaxies may be ideal laboratories to study massive bulge formation through pathways such as early gas-rich accretion, violent disk instabilities or misaligned infall of gas throughout cosmic time. 

\end{abstract}


\section{Introduction}
Galaxies with stellar masses comparable to the Milky Way ($M_* \sim 6 \times 10^{10}M_{\odot}$) –-- MW peers hereafter –-- host the majority of the stellar mass in the present-day Universe \citep{papovich15}. 
This mass range exhibits the greatest diversity in galaxy morphology and star formation activity for galaxies in the centers of their own halos, 
from nearly bulgeless star-forming galaxies like the Milky Way or M101 through to elliptical or lenticular galaxies like Centaurus A or the Sombrero. Furthermore, the mass of the bulge broadly correlates with black hole (BH) mass \citep{kormendy_ho}.
In turn, the BH mass strongly correlates with star formation activity –-- the more prominent the BH, the more quiescent the galaxy \citep{terrazas_pap1}. 

What drives the growth of MW peer bulges and BHs? It is often argued that massive, centrally concentrated ‘classical’ bulges are the result of major or minor merger activity whereas less massive, less centrally concentrated ‘pseudobulges’ may be the result of disk instabilities and bar evolution \citep[e.g.,][]{kormendy_kennicutt}.
Many galaxy formation models explicitly appeal to mergers to form bulges and BHs \citep[e.g.,][]{hopkins10,somerville15}. Indeed, it has been clear for decades that at least some elliptical galaxies are merger products \citep{toomre77,rothberg06}. Yet, other viable formation paths remain, e.g., early formation of a central bulge in a chaotic collapse \citep[e.g.,][]{johan12}, the formation of bulges through violent disk instabilities in early gas rich disks \citep[e.g.,][]{cev15}, or the formation of bulges from dramatic changes in the angular momentum of successive generations of accreted gas \citep{sales12}.
Indeed, given the active merging and accretion characteristic of a ΛCDM universe, and the relative infrequency of large bulges in the local Universe \citep{kormendy10} current generations of hydrodynamical models appear to dramatically over-produce bulges \citep[e.g.,][]{somerville15,brooks16}. While simulators explore possible processes to suppress bulge formation \citep[see, e.g.,][]{brooks16}, it is clear that observational insight into the relationship between galaxy merging and bulge growth is urgently required.

The outskirts of galaxies give particular insight into their merger histories. While gas in mergers can lose angular momentum and fall into the central parts of galaxies \citep[e.g.,][]{robertson05}, stars in merging galaxies are collisionless. Stars in low-mass satellite galaxies are easily tidally torn from their parent satellite and are spread out into a diffuse stellar halo \citep[e.g.,][]{bullock05,cooper10}. The stellar mass of galaxies is a strong function of dark matter halo mass; accordingly, the mass of the resulting stellar halo is dominated by the largest few disrupted satellites \citep[e.g.,][]{pur07}.
Since galaxies show a strong relationship between their metallicity and stellar mass \citep{kirby13}, stellar halos are predicted to and indeed show a strong correlation between their metallicity and mass \citep[\S \ref{sec:corr}]{deason16,harmsen16}. 
Merging with larger satellites may modify this trend by kicking up {\it in-situ} stars from the central galaxy {\it via} tidal tails or violent relaxation (e.g., \citealp{watkins15}). These stars will augment the overall population and will tend towards being metal rich, preserving the overall trend towards bigger satellites producing more massive and metal-rich stellar halos.

Thus, it is important to probe correlations between the stellar halo mass --- a sensitive record of galactic merger history --- and the prominence of a galactic bulge and its supermassive BH. Yet, stellar halo masses have been challenging to measure owing to their extremely low surface brightnesses, fainter than 30 $V$-band mag/arcsec$^2$. Recently, imaging of diffuse light from halos permitted detection of relatively massive halos \citep[e.g.,][]{Trujillo16,merritt16}. In parallel, observationally-expensive studies of resolved, typically red giant branch (RGB), stars in the outskirts of nearby galaxies reach fainter equivalent surface brightness limits and give estimates of both stellar halo masses and typical metallicities \citep{rs11,ibata14,rejkuba14,peacock15,mon16_ghosts,harmsen16} . 

The goal of this {\it Letter} is to use these recent measurements of stellar halo masses and metallicities as a new probe of the relationship between galactic merger histories, bulges and supermassive BHs. We focus on galaxies which are the main/central one in their dark matter halos with total stellar masses in the range $3-12\times 10^{10} M_{\odot}$. After reviewing observational estimates of stellar halo properties and bulge/BH masses in \S \ref{sec:halo} for a sample of 18 galaxies --- the only ones currently with suitable stellar halo measurements or limits --- we then show that there is little correlation between merger history and bulge or BH prominence (\S \ref{sec:corr}). We explore this issue further and review our main conclusions in \S \ref{sec:disc}.

\section{The Data}
\label{sec:halo}

We use estimates of total stellar halo mass and a representative stellar halo metallicity for this work. Stellar halo masses and metallicities are unavoidably ambiguous measurements --- both the bulge and stellar halo are kinematically hot and may be difficult to separate, and while the stellar halo is more likely to be accretion-dominated and the bulge {\it in situ}-dominated, both components are expected to have contributions from both accretion and {\it in situ} star formation \citep{zolotov09}. While this highlights the importance of fairly connecting simulations and observations, current efforts have not yet reached this goal. 
Therefore we take a simple approach that is easy for simulators to model. 

Observations typically measure the extended diffuse component at galactocentric minor axis distances between 10--40\,kpc and $>$20\,kpc along the major axis \citep{merritt16,harmsen16}. We extrapolate `aperture' stellar halo masses within this radial range to total stellar halo mass using accreted stars in cosmologically-motivated models. Such extrapolations are inherently uncertain and 
may vary with merger history or bulge prominence \citep{amorisco_atlas17,rodriguez-gomez16}. We explore this issue using accreted stellar halos of $>1200$ galaxies with $M_*$ between $3-10 \times 10^{10} M_{\odot}$ in the {\it Illustris} hydrodynamical model \citep{rodriguez-gomez16}. The total accreted stellar halo mass is $\sim 3 \times$ the `aperture' accreted mass with a scatter of 40\% and $<15$\% systematic variation across the range of halo masses and galaxy morphologies. These results agree quantitatively with our analysis of \citet{bullock05} models \citep{harmsen16} and preliminary analysis of the accreted halos of MW peers in the high-resolution {\it Auriga} hydrodymanical simulations (A.\ Monachesi, in prep.). We adopt this extrapolation and its uncertainties in what follows.    

As hydrodynamical simulations that have both accreted and {\it in situ} components find that minor axis metallicity profiles at galactocentric radii $>$15\,kpc are uncontaminated by {\it in situ} material \citep{mon16_sim}, we choose to estimate stellar halo metallicity at a {\it minor axis} distance of 30\,kpc (corresponding to $\sim 1/10$ of the virial radius).  

Six galaxies in our sample have resolved stellar populations information for their stellar halos from the GHOSTS (Galaxy Halos, Outer disks, Substructure, Thick disks and Star clusters) survey (\url{http://vo.aip.de/ghosts/survey.html}; \citealp{rs11}, \citealp{mon16_ghosts}). RGB star counts allowed GHOSTS to reach $V$-band (Vega) surface brightness sensitivities of $\sim$33 mag/arcsec$^2$, revealing roughly power-law stellar halos between minor axis distances of 10\,kpc to $>$50\,kpc.
These profiles were integrated within the observed range of radii and extrapolated to total stellar halo mass as discussed above. 
The RGB color is sensitive to stellar halo metallicity \citep{mon16_ghosts}, and we adopt a derived [Fe/H] value at 30\,kpc along the minor axis following the observational calibration of [Fe/H] as a function of RGB colors for globular clusters \citep{streich14} assuming $[\alpha/{\rm Fe}] =0.3$.

We add stellar halo masses and metallicities at 30\,kpc along the minor axis for the MW \citep{bh16,sesar11,xue15}, M31 \citep{ibata14,gilbert14} and NGC\,3115 \citep{peacock15}. \citet{peacock15} estimated a mass in low metallicity stars in NGC\,3115 of $\sim 1.5\times 10^{10}M_{\odot}$, which they adopted as an estimate of the stellar halo mass. As there is a considerable population of high metallicity stars at large radii in NGC\,3115 which could be argued to legitimately belong to its stellar halo, we choose to adopt $1.5\times 10^{10}M_{\odot}$ as a lower limit to its halo mass. No published estimate of stellar halo mass exists for Cen A; we adopt the metallicity along the minor axis at 30\,kpc from \citet{rejkuba14}.   

\begin{deluxetable*}{ccccccc}

%% Keep a portrait orientation
%% Over-ride the default font size
%% Use 10pt
\tabletypesize{\footnotesize}

%% Use \tablewidth{?pt} to over-ride the default table width.
%% If you are unhappy with the default look at the end of the
%% *.log file to see what the default was set at before adjusting
%% this value.

\tablecaption{Galaxy Properties \label{tab:data}} 

%% The \tablehead gives provides the column headers.  It
%% is currently set up so that the column labels are on the
%% top line and the units surrounded by ()s are in the 
%% bottom line.  You may add more header information by writing
%% another line between these lines. For each column that requries
%% extra information be sure to include a \colhead{text} command
%% and remember to end any extra lines with \\ and include the 
%% correct number of &s.
\tablehead{\colhead{Name} & \colhead{Stellar Mass} & \colhead{Stellar Halo Mass} & \colhead{Stellar Halo [Fe/H]} & \colhead{B/T} & \colhead{Black Hole Mass} & \colhead{References} \\ 
\colhead{} & \colhead{($10^{10} M_{\odot}$)} & \colhead{($10^{9} M_{\odot}$)} & \colhead{(dex)} & \colhead{} & \colhead{($10^{6} M_{\odot}$)} & \colhead{} } 
%% All data must appear between the \startdata and \enddata commands
\startdata
Milky Way &  $6.1$ &  $0.55\pm0.15$ &  $-1.7 \pm 0.1$ &  $0.2\pm0.075$ &  $4.3\pm0.36$ &  1;2;3;4;5 \\
M31 &  $10.3$ &  $15\pm5$ &  $-0.75\pm0.2$ &  $0.32\pm0.11$$^{*}$ &  $143^{+91}_{-31}$ &  6;7;8;9;10;5 \\
NGC 253 &  $5.5$ &  $4.5\pm2$ &  $-1.05\pm0.1$ &  $0.28\pm0.14$ &  $<7$ &  11;8;12;13 \\
NGC 891 &  $5.3$ &  $2.7\pm1.1$ &  $-1.0\pm0.1$ &  $0.2\pm0.05$ &  \nodata &  11;8;14 \\
M81 &  $5.6$ &  $1.1\pm0.5$ &  $-1.3\pm0.1$ &  $0.46\pm0.15$$^{*}$ &  $65^{+25}_{-15}$ &  11;8;12;5 \\
NGC 4565 &  $8.0$ &  $2.2\pm0.9$ &  $-1.2\pm0.1$ &  $0.25\pm0.05$ &  \nodata &  11;8;15 \\
NGC 4945 &  $3.8$ &  $3.5\pm1.4$ &  $-0.85\pm0.1$ &  $0.075\pm0.025$ &  $1.35^{+0.48}_{-0.68}$ &  11;8;5 \\
NGC 7814 &  $4.5$ &  $6.4\pm2.6$ &  $-1.05\pm0.1$ &  $0.85\pm0.15$$^{*}$ &  \nodata &  11;8;12 \\
Cen A &  $11.2$ &  \nodata &  $-0.43\pm0.1$ &  $1^{+0.0}_{-0.1}$$^{*}$ &  $57\pm10$ &  5;16 \\
NGC 3115 &  $10.5$ &  $>15$ &  $-0.6\pm0.2$ &  $0.8\pm0.1$$^{*}$ &  $897^{+53}_{-277}$ &  5;17 \\
UGC 180 &  $13.0$ &  $4.0\pm2.4$ &  \nodata &  $0.1\pm0.05$ &  \nodata &  18 \\
NGC 1084 &  $4.3$ &  $6.3\pm3.0$ &  \nodata &  $0.04\pm0.02$ &  \nodata &  19;11;12 \\
NGC 2903 &  $4.9$ &  $1.5\pm1.0$ &  \nodata &  $0.07\pm0.03$ &  \nodata &  19;11;12 \\
NGC 3351 &  $5.8$ &  $<4.0$ &  \nodata &  $0.15\pm0.07$ &  $<9.7$ &  19;11;20;12 \\
NGC 3368 &  $8.9$ &  $<8.8$ &  \nodata &  $0.32\pm0.05$ &  $7.7^{+1.5}_{-1.6}$ &  19;11;21;5 \\
NGC 4220 &  $6.1$ &  $1.6\pm1.3$ &  \nodata &  $0.12\pm0.06$ &  \nodata &  19;11;12 \\
NGC 4258 &  $7.6$ &  $<4.3$ &  \nodata &  $0.12\pm0.03$$^{*}$ &  $37.8\pm0.4$ &  19;11;22;5 \\
M101 &  $5.9$ &  $<0.7$ &  \nodata &  $0.05\pm0.03$ &  $<3$ &  19;11;12;22 \\
\enddata

%% Include any \tablenotetext{key}{text}, \tablerefs{ref list},
%% or \tablecomments{text} between the \enddata and 
%% \end{deluxetable} commands

\tablecomments{Stellar masses are assumed to have an uncertainty of $0.15$ dex. \\
$^{*}$ --- Galaxies with dominant classical bulges. }

%% General table references marker
\tablerefs{1 -- \cite{licquia15}; 2 -- \cite{bh16}; 3 -- \cite{sesar11}; 4 -- \cite{xue15}; 5 -- \cite{kormendy_ho}; 6 -- \cite{sick15}; 7 -- \cite{ibata14}; 8 -- \cite{mon16_ghosts}; 9 -- \cite{gilbert14}; 10 -- \cite{tamm12}; 11 -- \cite{harmsen16}; 12 -- \cite{salo15}; 13 -- \cite{rodriguez06}; 14 -- \cite{schecht13}; 15 -- \cite{schecht14}; 16 -- \cite{rejkuba14}; 17 -- \cite{peacock15}; 18 -- \cite{Trujillo16}; 19 -- \cite{merritt16}; 20 -- \cite{sarzi02}; 21 -- \cite{mollenhoff01}; 22 -- \cite{kormendy10} }

\end{deluxetable*}

Stellar halo masses (for UGC 180, NGC 1084, NGC 2903 and NGC 4220) or upper limits (for NGC 3351, NGC 3368, NGC 4258 and M101; galaxies that were either presented as upper limits or detections with 1$\sigma$ error bars overlapping with zero) are adopted from \citet[UGC 180]{Trujillo16} and \citet{merritt16}. The stellar halo masses from \citet{merritt16} are extrapolated to total stellar mass as described above and as 
tabulated by \citet{harmsen16}. As these halo masses were constrained using integrated light, these galaxies have no stellar halo metallicity estimates. 

Total stellar masses were adopted from the source papers, adjusted to a universally-applicable \cite{chabrier} stellar IMF. 
Bulge-to-total ratios were derived from near-IR (K-band or longer) photometry from a variety of sources (described in Table \ref{tab:data}; for Cen A we assume $B/T=1$ and UGC 180's $B/T$ was roughly estimated to be $\sim 0.1 \pm 0.05$). Typical differences between independent estimates for the same galaxies are $\Delta(B/T)\sim 0.1$, comparable to the quoted uncertainties. BH masses and uncertainties (or limits) were adopted primarily from \citet{kormendy_ho}, where again the source papers are tabulated in Table \ref{tab:data}. 

\section{Little correlation between stellar halo, bulge and black hole properties } \label{sec:corr}

\begin{figure}\centering
	\includegraphics[width=88mm]{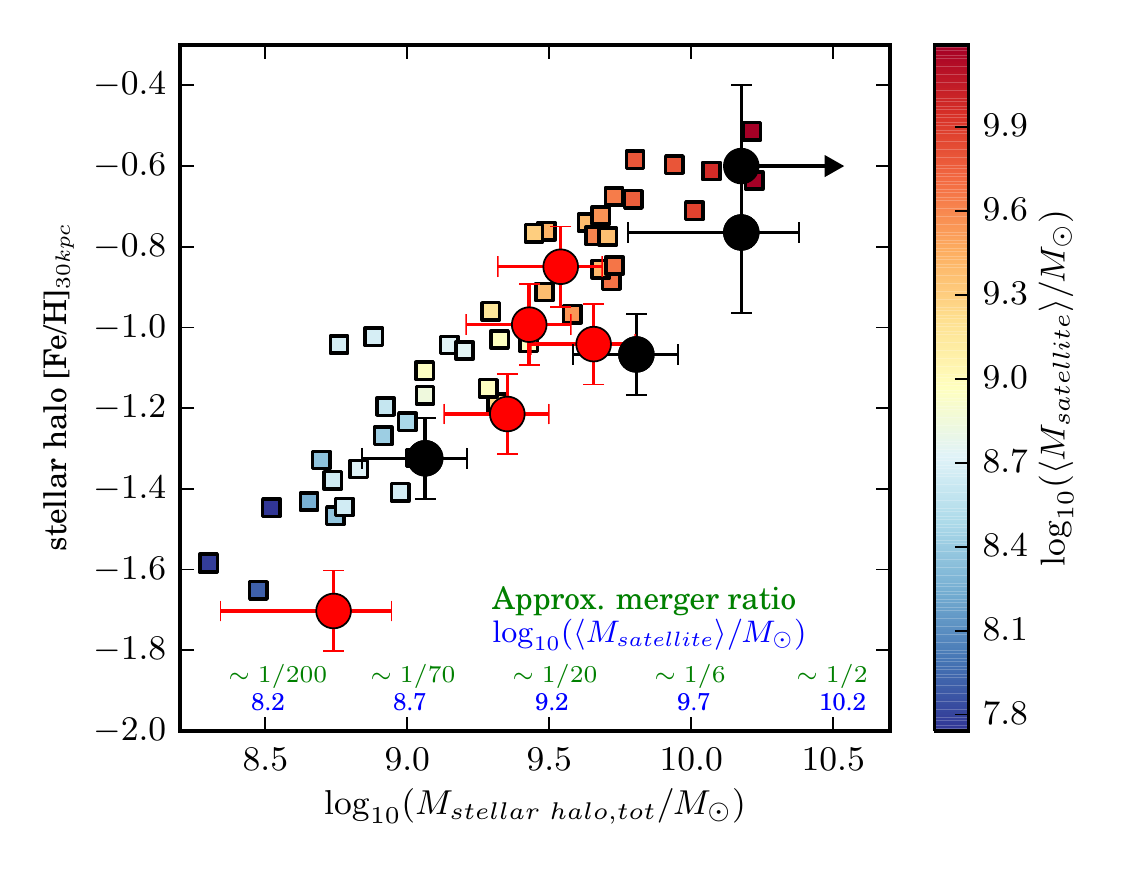}
	\caption{Circles show the observed correlation between stellar halo mass derived from integrating star counts and stellar halo metallicity derived from minor axis RGB color at 30kpc. Galaxies with ‘classical’ bulges are shown in black, galaxies with ‘pseudobulges’ are shown in red. Squares show {\it simulated} stellar halo masses and metallicities from the accretion-only models of \citet{deason16}. The squares are color-coded according to the {\it mass weighted} mean stellar mass of the contributing satellites to the halo; we indicate the approximate run of mass weighted mean accreted satellite mass in blue and corresponding approximate merger ratios for a main galaxy mass at the time of merger of $\sim 3\times10^{10}M_{\odot}$ in green.  
}
    \label{fig:metmass}
\end{figure}

\begin{figure*}[t]\centering
	\includegraphics[width=135mm]{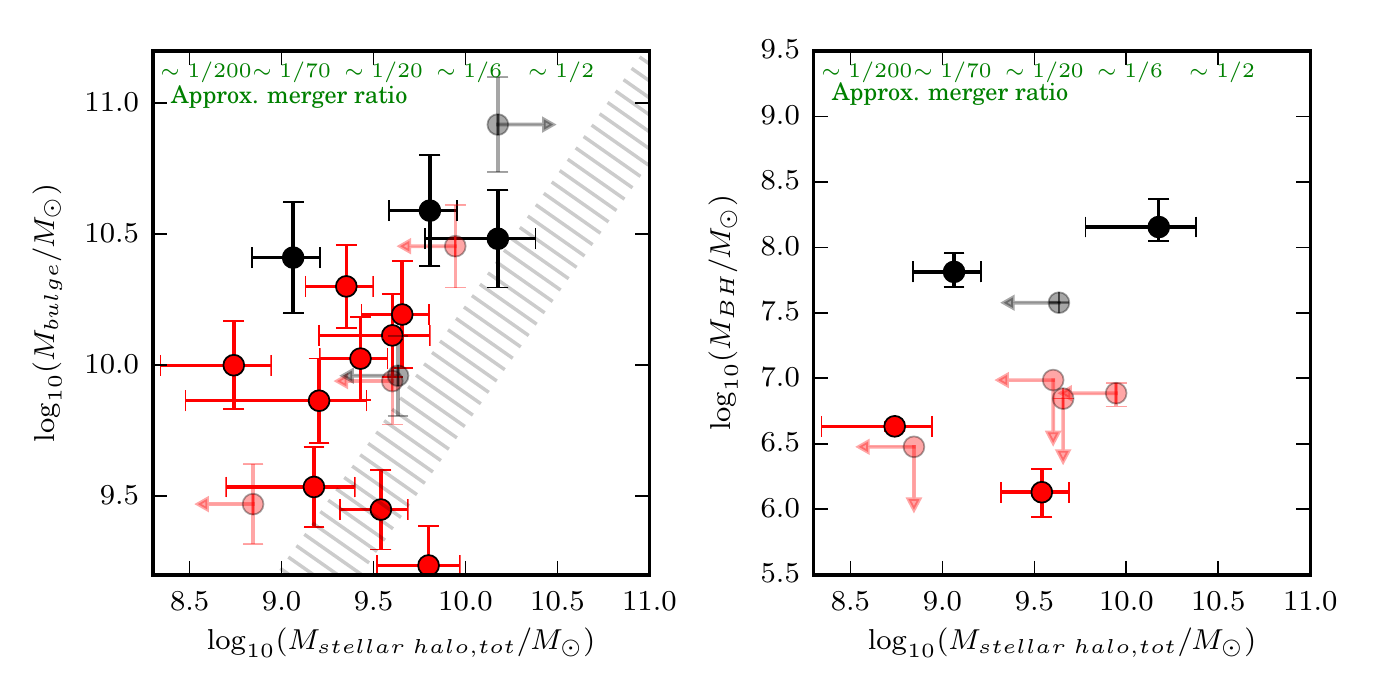}
	\caption{Bulge mass (left) and BH mass (right) as a function of stellar halo mass. Red denotes galaxies with low-mass ‘pseudobulges’, black shows galaxies with higher-mass ‘classical’ bulges; limits are shown with lighter shading. As argued in Fig.\ \protect\ref{fig:metmass}, stellar halo mass reflects merger history and we give approximate merger ratios in green. The shaded grey area in the left panel schematically illustrates what would be expected if there were a 1:1 correlation between stellar halo mass and bulge mass, as broadly expected in some simple modeling contexts (e.g., \protect\citealp{hopkins10}). 
}
    \label{fig:corr}
\end{figure*}

\begin{figure}[t]\centering
	\includegraphics[width=75mm]{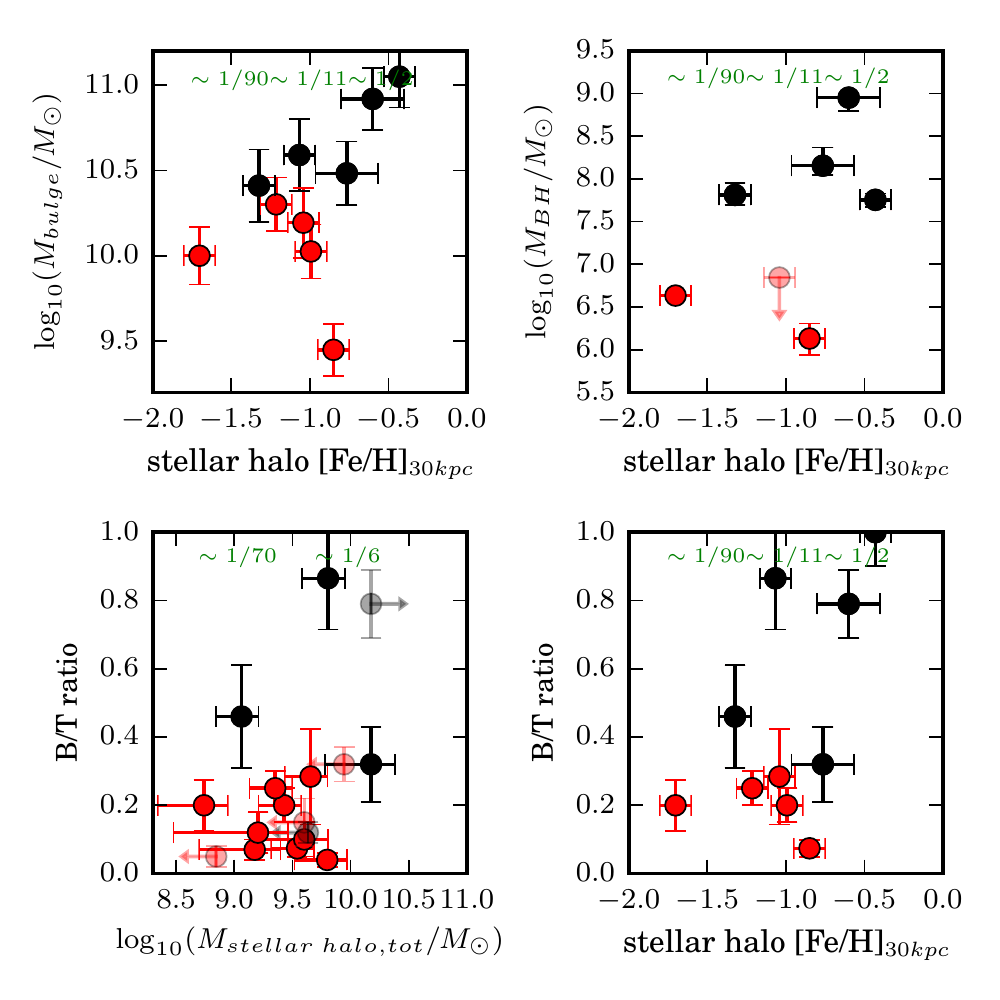}
	\caption{Bulge mass (top left) and BH mass (top right) as a function of stellar halo metallicity. The bottom panels show the B/T ratio as a function of stellar halo mass (left) and metallicity (right). Red denotes galaxies with low-mass ‘pseudobulges’, black shows galaxies with higher-mass ‘classical’ bulges; limits are shown with lighter shading. We give approximate merger ratios in green.  
}
    \label{fig:corr2}
\end{figure}

One of the foundational assumptions of this work is that the stellar halo mass or metallicity is a reliable record of the merger history of a galaxy. We examine this assumption in Fig.\ \ref{fig:metmass}. There is a strong correlation between the observed stellar halo masses and their metallicities at 30\,kpc along the minor axis \citep{harmsen16}. This correlation is a generic prediction of accretion-only models, driven by the metallicity-mass relation of the disrupted satellite galaxies \citep{deason16,harmsen16}. Quantitative insight can be gained by comparison with a set of modeled stellar halos in $10^{12} M_{\odot}$ dark matter halos from \citet[squares]{deason16}. These models are color-coded by the mass weighted mean stellar mass of all of the contributing satellites to the halo (termed `typical' satellite mass hereafter), illustrating that the stars in more massive stellar halos primarily hail from more massive, more metal-rich satellite galaxies. The `typical' satellite mass is a tight function of the stellar halo mass and we provide an approximate mapping on the ordinate in blue. We give also an  approximate merger ratio assuming a main galaxy mass at the time of merger of $\sim 3 \times 10^{10} M_{\odot}$ in green. 
The observation of this expected correlation strongly suggests that indeed the stellar halo encodes the merger history of galaxies, and that we can now broadly infer the masses of the largest satellites that were accreted by or merged into our nearby neighbors. 

We can now explore how MW peer merger history correlates with bulge and BH prominence. The main results are shown in Figure \ref{fig:corr} where we show the bulge mass (left) and BH mass (right) as a function of stellar halo mass. For completeness, we show bulge mass and BH mass as a function of stellar halo metallicity in the top panels of Fig.\ \ref{fig:corr2}, and B/T ratio in the lower panels as a function of stellar halo mass (left) and metallicity (right). Both stellar halo mass and metallicity reflect the merger history of a galaxy as galaxies with bigger stellar halos merged with/accreted larger progenitors. {\it No significant correlations between bulge/BH masses and stellar halo masses are seen in this dataset.} Galaxies with $M_{stellar\,halo,tot}>10^9 M_{\odot}$ or [Fe/H]$_{30\,kpc} > -1.2$ have an order of magnitude spread in B/T ratio or bulge mass and two orders of magnitude spread in BH mass. All of the relations have Spearman rank correlation coefficients consistent with being drawn from uncorrelated datasets at least 15\% of the time. 

There are six `classical' bulges in our sample, shown in black, all but one (NGC 4258) with masses in excess of $2 \times 10^{10} M_{\odot}$ –-- these could plausibly have been expected to have been formed in (minor or major) mergers \citep[e.g.,][]{kormendy_kennicutt}. Three galaxies with classical bulges (M31, NGC 3115, Cen A) indeed have massive metal-rich stellar halos --- carrying $>20$\% of the total galaxy stellar mass --- indicative of a minor or major merger. Three (M81, NGC 4258, NGC 7814) have less massive stellar halos. NGC 4258 has a low mass classical bulge and a relatively uninformative limit on its stellar halo mass. NGC 7814 has a very large bulge five times more massive than its stellar halo. Most notable among these is M81, with a large classical bulge and an anemic stellar halo containing only $2\pm0.9$\% of its total stellar mass. M81 shows {\it no sign of any significant past major or minor merging activity that was expected to drive the formation of its classical bulge}.

\section{Discussion and conclusions}
\label{sec:disc}

MW peer galaxies have a wide range in stellar halo masses and therefore merger histories which appear not to correlate significantly with bulge to total ratio, bulge mass or BH mass. This appears to pose a challenge for models where galaxy bulges (even just classical bulges) are formed primarily by minor or major merging activity.

A detailed comparison with hydrodynamical models is desirable but beyond the scope of this {\it Letter}. A principled and informative comparison would have high enough resolution to model bulge growth and stellar feedback realistically in the main and satellite galaxies, a large enough dynamic range and sample size to model a range of merger histories and halo masses, and include an effort to connect observational and simulated metrics fairly. Such simulations are coming online soon and should shed light on this issue. 

In the meantime, qualitative insight can be gained by comparing with trends predicted by simplified models incorporating a wide range of binary mergers. \citet{hopkins10} expects $B/T \sim M_{\rm secondary}/M_{\rm primary}$ for star-dominated progenitors; such a trend is qualitatively illustrated in the center left panel of Fig.\ \ref{fig:corr} by assuming $M_{\rm stellar\ halo} \sim M_{\rm secondary}$, giving $M_{\rm bulge}\sim M_{\rm stellar\ halo}$. The observations do not follow such a trend. Interestingly, observed bulges are more massive (sometimes by more than an order of magnitude) than their stellar halos, implying substantial growth of bulges not directly connected to mergers with star-rich satellites. 

Much remains to be learned about the use of stellar halo masses and metallicities as metrics of merger history. While relatively late star-rich mergers inevitably leave debris at large radii in an extended halo \citep{bullock05,cooper10}, early gas-rich mergers could leave little in the way of stellar halo signature \citep{robertson05}, or leave a signature consistent with an {\it in situ} stellar halo \citep{cooper15}. This issue should be able to be explored with the upcoming suites of high-resolution simulations. 
Our observations nonetheless demonstrate that there is a poor correlation between merger history of star-rich progenitors/satellites (likely $z<2$) and the bulge/BH prominence.

A further interpretive challenge is highlighted by M81. \citet{okamoto15} show that M81 is currently interacting with M82 and NGC 3077;  the future disruption of these satellites and their effects on M81 will likely form a large, metal-rich stellar halo. This interaction may also enhance M81's future bulge and black hole. Yet, given that M81 is already $\sim 1/2$ bulge, M81's future stellar halo will have little to do with the process that created most of M81's bulge and massive BH. Accordingly, while a large metal-rich halo implies a significant minor or major merging event, this does not necessarily imply that the bulge was formed as a result of this event (correlation does not imply causation). 

It may be tempting to appeal to early bulge formation via processes unrelated to merging at e.g., $z>2$. Yet, \citet{vd13} find that bulges emerge continually in progenitors of MW peers, with much of the bulge mass coming into place at $z<1$. Furthermore, given that $z<2$ merging of star-rich progenitors occurs and will contribute to bulges, an excessive amount of early bulge production will dramatically over-produce bulges in the local Universe \citep{brooks16}.

Putting this together, we suggest the following interpretation of our results.
\begin{enumerate}
\item A wide range of (quieter) merger histories, giving an order of magnitude range in stellar halo masses from $\sim 3\times 10^8 M_{\odot}$ to $\sim 3\times 10^9 M_{\odot}$, result in galaxies with small bulges with `pseudobulge'-like properties (highlighted in red in Figs.\ \ref{fig:metmass}--\ref{fig:corr2}). Low mass bulge formation mechanisms that do not require intense merging activity (such as secular evolution, disk instabilities or misaligned gas accretion) appear at first sight to be consistent with our observations for these galaxies.
\item Similarly, a wide range of merger histories can give rise to massive, `classical' bulges with massive central BHs (denoted in black in Figs.\ \ref{fig:metmass}--\ref{fig:corr2}). Three out of five local galaxies (M31, NGC 3115 and Cen A) with massive classical bulges have massive stellar halos $M_{\rm halo} > 2 \times 10^{10} M_{\odot}$ consistent with (but not requiring) a star-rich minor or major merger origin. The other galaxies with massive classical bulges appear to have lower mass stellar halos implying a quieter accretion history. These galaxies --- exemplified most vividly by M81 --- may be ideal laboratories to study massive bulge formation through pathways such as early gas-rich accretion, violent disk instabilities or misaligned infall of gas throughout cosmic time. 
\end{enumerate}

\acknowledgments
We thank the referee for their constructive suggestions. We thank Alis Deason for providing the properties of her modeled stellar halos in electronic form. We acknowledge useful discussions with Denija Crnojevic, Duncan Forbes, Sarah Loebman, Ian Roederer, Bryan Terrazas, \& Monica Valluri. This work was partially supported by NSF grant AST 1514835. This research has made use of NASA's Astrophysics Data System Bibliographic Services.

\end{document}